\documentclass[aps,prb,reprint,groupedaddress,showpacs]{revtex4-1}
\usepackage{amssymb}
\usepackage{graphicx}

\begin{document}

\title{Dynamics of superconducting vortices driven by oscillatory forces in the plastic flow regime}

\author{D. P\'{e}rez Daroca}
\email{daroca@df.uba.ar}
\affiliation{Departamento de F\'{\i}sica, FCEyN, Universidad de Buenos Aires and IFIBA, 
CONICET; Pabellon 1, Ciudad Universitaria, 1428 Buenos Aires, Argentina}
\author{G. Pasquini}
\affiliation{Departamento de F\'{\i}sica, FCEyN, Universidad de Buenos Aires and IFIBA,
CONICET; Pabellon 1, Ciudad Universitaria, 1428 Buenos Aires, Argentina}
\author{G. S. Lozano}
\affiliation{Departamento de F\'{\i}sica, FCEyN, Universidad de Buenos Aires and IFIBA,
CONICET; Pabellon 1, Ciudad Universitaria, 1428 Buenos Aires, Argentina}
\author{V. Bekeris}
\affiliation{Departamento de F\'{\i}sica, FCEyN, Universidad de Buenos Aires and IFIBA,
CONICET; Pabellon 1, Ciudad Universitaria, 1428 Buenos Aires, Argentina}

\begin{abstract}

We study experimentally and theoretically, the reorganization of superconducting   vortices driven by oscillatory forces near the plastic depinning transition.  We show that the system can be taken to configurations that are tagged by the shaking parameters but keep no trace of the initial conditions. In experiments performed in $NbSe_2$ crystals, the periodic drive is induced by ac magnetic shaking fields and the overall order of the resulting configuration is determined by non invasive ac susceptibility measurements. With a model of interacting particles driven over random landscapes, we perform molecular dynamics simulations that reveal the nature of the shaking dynamics as fluctuating states similar to those predicted for other interacting particle systems.
\end{abstract}

\maketitle

Vortices  in type II superconductors \cite{vortex, vortex2, giamarchi95,koshelev,pippard,andrei,Zeldov,xiao,pasquini08,diego, reichhardt06}, charge density waves \cite{charge density},
 colloidal particles \cite{colloidal,reichhardt, nature2008}, Wigner crystals \cite{Wigner crystals} and
 domain walls \cite{domain} are only few  of the many examples of physical
 systems belonging to the category of driven
elastic manifolds over random landscapes. In all these systems, competing
interactions give rise to a complex dynamics characterized by a region of
plastic motion: under the action of an external driving force, strong enough
to depin the system, a non-linear response develops, some particles are
mobile while others remain pinned. This plastic depinning and its
relation with the proliferation of topological defects has been object of
study for several years. Recently, it has been claimed \cite{reichhardt}
that the plastic depinning of a colloidal system with
random quenched disorder under the action of a dc force can be described in
terms of a non equilibrium phase transition with a divergent transient time
and it can be related to the "absorbing transition" from a random self-organization to a drive-dependent
fluctuating steady state (FSS) observed in experiments on
periodically sheared particle suspensions \cite{nature2008}.

Vortex matter offers an ideal playground to study this phenomenology both numerically and
experimentally. Dynamic phase transition \cite{vortex,koshelev} and plastic flow have been theoretically described and
measured in the neighborhood of the order-disorder (O-D) transition,
where the system evolves from a quasi-ordered Bragg glass (BG)
\cite{giamarchi95} to a disordered phase with proliferation of
topological defects. The experimental fingerprint of the O-D
transition is the anomalous non-monotonous dependence of the
critical current density $J_{_{C}}$ with both temperature and
magnetic field, known as Peak Effect (PE) \cite{pippard}. There,
 the response at a fixed temperature and field can depend on
thermal, magnetic and dynamical history that can modify
 the topology of the vortex lattice configurations (VLCs)
by creation or annihilation of disclinations \cite{andrei}. In
transport experiments vortex instability phenomena and the smearing of
the PE have been claimed to originate on a competition between the
injection of a disordered vortex phase at the sample edges, and the
dynamic annealing of this metastable disorder by the transport
current \cite{Zeldov}. However, using an  ultrafast 
transport technique that avoids current-induced vortex lattice (VL)
reorganization, Xiao \emph{et al.} \cite{xiao} have shown
the existence of an enlarged crossing phase boundary between ordered and disordered phases in
$NbSe_{2}$ crystals arising from
the bulk VL response, that has been corroborated in a recent work by
our group using non invasive ac susceptibility measurements in the
linear regime\cite{pasquini08}. In experiments, a shaking ac field
is often applied to order the VL and the most accepted picture is
that the shaking field assists the system in an equilibration
process, from a disordered metastable configuration to the
equilibrium BG phase. However, in our experiments, we have shown
that in the transition region a\ shaking ac field can either order
or disorder the VL, driving the system to VLCs with
intermediate degree of disorder that are independent of the initial
VLC. Very recently, noise transport experiments (in a-$\text{Mo}_x\text{Ge}_{1-x}$) in this 
intermediate region have shown evidence of a depinning transition with critical
 behavior similar to the absorbing transition \cite{vortex2}. 

In this work  we study experimentally and numerically the
 reorganization of a vortex system in the plastic flow region under the action
of ac shaking magnetic fields. We have done experiments changing the frequency and
the waveform of the shaking fields and measuring the ac response of the final
VLC in each case with very low ac amplitude, avoiding modifications in the final VLC. We have also performed numerical calculations that mimic our experiment. We
have first used the Brandt method \cite{brandt} to determine the induced current
dependence on the external magnetic field and as a second step we have used
this result as input for a molecular dynamics simulation. Although the model
is rather simple, we obtain results which are qualitatively consistent with
the experimental results and show evidence of ac driven vortex lattice reorganization.

Experimental results shown in this work correspond to a $NbSe_{2}$ single
crystal \cite{crystal, similar} of approximate dimensions $(0.5\times 0.5\times
0.03)\ mm^{3}$, with $Tc\,=\,7.30\ K$ (defined as the mid point of the ac
susceptibility linear transition at $H=0$) and $\Delta Tc=\pm \,0.02\,K$. 
A neutron study \cite{yaron} in samples synthesized with the same technique as
ours, shows an excellent agreement with the VL structure predicted
 for randomly distributed point defects \cite{lando}.  The ac susceptibility has been measured with
a home made susceptometer based in the mutual inductance technique, where
both the ac and dc fields are parallel to the $c$ axis of the sample. To
study the VL response without disturbances, the measurements have been
restricted to the linear Campbell regime \cite{campbell}, where a very small ac field $h_{a}$
superimposed to the dc field $H$ is applied, forcing vortices to perform
small (harmonic) oscillations inside their effective pinning potential
wells. In this regime, the ac susceptibility is independent of amplitude and frequency and the inductive component of the ac susceptibility $\chi
^{,}$ is determined by the experimental geometry and the curvature of the
effective pinning potential well $\alpha _{L}$ and a lower
susceptibility $\chi ^{,}$ (closer to -1) can be directly associated to a
more pinned VL  (see Ref. \onlinecite{gabi99}). Therefore, at
fixed $T$ and $H$, different $\chi ^{,}$ (i.\thinspace e. different degree
of effective pinning) can be
associated to the existence of VLCs with different degree of disorder.

\begin{figure}[t]
\includegraphics[width=86mm]{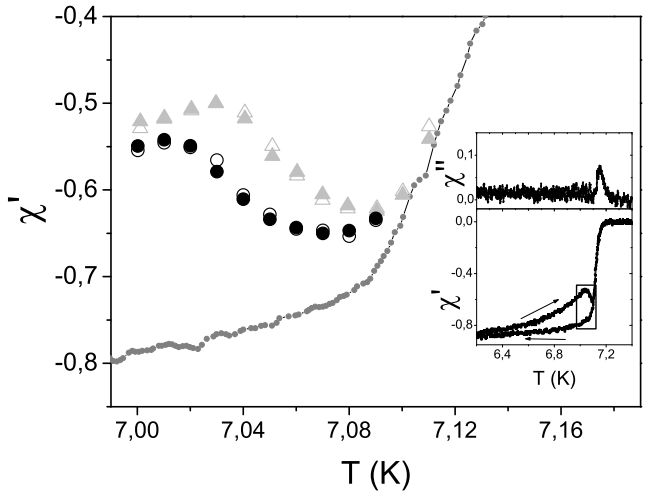}
\caption{(Inset) Typical $\protect\chi ^{,,}$ and $\protect\chi ^{,}$ FCC
and FCW curves in the linear regime measured at $f\,=\,30\,kHz$. Arrows indicate the direction 
of temperature variation. (Main panel) Zoom of the
region delimited in the inset. Gray triangles (black circles) correspond to  
 shaking frequency $f_{sh}\,=\,100\,kHz$ ($%
f_{sh}\,=\,10\,Hz$). Open (full) symbols for warming (cooling) experiment.
 In darkgray small circles a FCC curve is shown.}
\label{fig1}
\end{figure}

Figure \ref{fig1} shows $\chi ^{,}(T)$ measured in the linear regime ($h_{a}\,=\,0.025\,Oe$,
$f\,=\,30\,kHz$), in a dc field $H\,=\,320\,Oe$ after different thermal and
dynamical histories. In the Inset, both components $\chi ^{,}$ and $\chi
^{,,}$ in field cool cooling (FCC) and field cool warming (FCW) processes
are shown in the full temperature range, and the region of the PE (main
panel) is indicated. The curve with small darkgray circles in the main panel was obtained in a
FCC process, where the VL remains trapped in a metastable disordered and
strong pinned configuration, and the PE disappears. The other experimental
points have been obtained by measuring the system after shaking the VL \ at
different frequencies. In all those cases, the following procedure has been
performed: The system has been stabilized at a shaking temperature $T_{sh}$
and measured in the linear regime, recording the $\chi ^{,}$ value corresponding to the starting configuration.
 Then the measurement
was interrupted and $N_{sh}$ cycles of a sinusoidal shaking ac field ($%
h_{sh}\,=\,3.2\,Oe$) were applied. After switching off the shaking
field, $\chi ^{,}$ corresponding to the final value was recorded in
the linear regime; then the system was driven to the next shaking
temperature. Full and open symbols correspond to procedures
increasing and decreasing $T_{sh}$. For clarity, in this figure only
the final values are shown. Gray triangles indicate the response after
shaking the system with $N_{sh}=100$ cycles at a frequency $%
f_{sh}\,=\,100\,kHz$, whereas black circles show the  response after a
similar procedure at $f_{sh}\,=\,10\,Hz$. One remarkable feature
published in our previous work \cite{pasquini08} is that the final VLCs do not show appreciable relaxation (within our experimental time window) and are independent on the initial condition. Surprisingly, we find that these final states are not unique: they depend on the shaking frequency, but however the independence of initial conditions persists.

\begin{figure}[t]
\includegraphics[width=86mm]{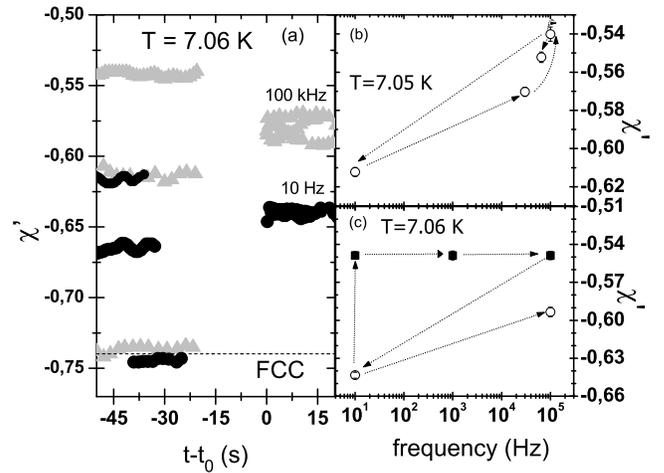}
\caption{(a) $\protect\chi ^{,}$ versus time  before and after shaking the system (the
shaking time is not in scale and it is indicated by a gap without data).
Gray (black) symbol correspond to $f_{sh}\,=\,100\,kHz$ ($%
f_{sh}\,=\,10\,Hz$ ). (b) $\protect\chi ^{,}$ values versus
the shaking frequency (log scale). Arrows point from the values of $\chi^{,}$ before shaking to the
values obtained after shaking. (c) $\protect\chi ^{,}$ values versus
 the fundamental shaking frequency for sinusoidal (open dots) and
square (black squares) shaking waveforms.}
\label{fig2}
\end{figure}

In Fig. \ref{fig2}(a), the response at fixed temperature, before and after applying shaking protocols, is shown for different starting VLCs (same frequencies as in Fig. \ref{fig1}). It can be seen that the shaking field can order or disorder the VL. The various curves $\chi ^{,}(t)$ obtained at
different initial configurations collapse in an unique curve after shaking
the system at a given frequency. Figure \ref{fig2}(b) shows the response
after applying sinusoidal shaking fields at various frequencies.
 Arrows point from each starting
VLC to the corresponding final configuration after shaking. In all the
cases the final response is independent of the initial one, for shaking amplitudes greater than 1.8 Oe and above a number of applied cycles (in our system
$N_{sh} \gtrapprox $ $40$). For all the tested frequencies (not all shown in the figure), the higher the
shaking frequency, the more ordered the resulting VLC, consistent with a
dynamic reordering of the VL \cite{reichhardt06}.

In that scenario, the frequency dependence would be due mainly
 to the different vortex velocities (i.\,e. forces) induced by 
the different shaking ac fields. Additional
evidence supporting this fact arises from the response of
 the system to shaking fields with a square
waveform.  In this case it is expected that  vortices move at higher velocities during short time intervals triggered and determined by the rising and falling edges, whereas the shaking frequency only modifies the waiting time between these short intervals. In Fig  \ref{fig2}(c), it is seen that for square shaking fields,
the VLCs are maximally ordered.
On the other hand, frequency independence shows that
relaxation processes are not relevant at least in this time window.
As another evidence, we have confirmed that triangular and sinusoidal shaking fields (that at the same fundamental frequency induce similar vortex velocities) produce similar results.

\begin{figure}[t]
\includegraphics[width=86mm]{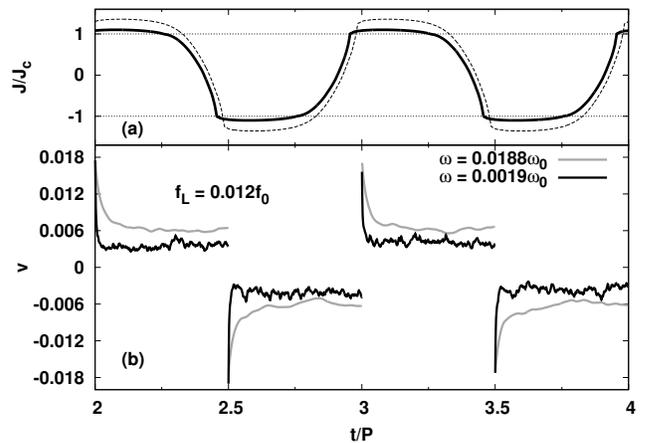}
\caption{(a) Theoretical current density $J/J_c$ at a fixed
position, in a superconducting disk of radius $a$ and a perpendicular sinusoidal magnetic field, as a function of $t/P$ for two different frequencies, 1$\tilde{\protect%
\omega}$, solid line and 100$\tilde{\protect\omega}$, dashed line, $\tilde{%
\protect\omega} = \frac{2 \protect\pi E_c}{\protect\mu_0 a J_c}$. Throughout most of the sample, the current can be
described by a square waveform. (b) Calculated vortex mean velocity v,
obtained from molecular dynamic simulations, as a function of $t/P$.}
\label{fig3}
\end{figure}

 The simulated system describes a very small region of the real
macroscopic sample, where the mean vortex density and mean 
current density can be assumed to be homogeneous. In the experiment, the
shaking ac field induces macroscopic currents $J_{sh}(t)$ that move
vortices, that have been introduced in our simulations by a Lorentz
force term $F^{L}=\phi _{0}J_{sh}\times \widehat{z}$, where
$\widehat{z}$ is the versor parallel to the vortex direction. The
explicit functional relation between $J_{sh}(t)$ and the shaking
$h_{ac}(t)$ is not straightforward. In order to gain some intuition, we have applied the method developed
in Ref. \onlinecite{brandt} to obtain the distribution of $J(t)$ for a certain geometry, assuming a non linear electric field-current constitutive relation  $E(J)=E_{c}(J/J_{c})^{n}$ \cite{nota}. We have chosen a simple
geometry (a disk) and we have estimated $n$ by comparing non linear
susceptibility measurements and polar plots of $\chi ^{,,}$ vs $\chi ^{,}$.
Under this non linear regime, given an external $h_{sh}(t)=h_{sh}\cos
(\omega t)$ the current throughout most of the sample  can be described by a
square waveform of frequency $\omega $ and normalized amplitude $
J_{0}(\omega )/J_{c}\gtrsim 1$ (Fig. \ref{fig3}(a)), being the frequency dependence
logarithmic. Thus as a first (and very crude) approximation we will simulate
our model with $J_{sh}(t)=J_{0}(\omega )\mbox{Sign}(\sin (\omega t))$.

In our simulations, we consider $N_{v}$ rigid vortices with coordinates $%
\mathbf{r}_{i}$ in a two-dimensional rectangle of size $L_{x}\times L_{y}$
that evolve according to the dynamics $\eta \mathbf{v}_{i}=\mathbf{F}%
_{i}^{vv}+\mathbf{F}_{i}^{vp}+\mathbf{F}^{L},\label{equ1 copy(1)}$ where $%
\mathbf{v}_{i}$ its velocity, $\eta $ the Bardeen-Stephen viscosity
coefficient, $\mathbf{F}_{i}^{vv}=\sum_{j%
\neq i}^{N_{v}} f_0 f_{vv}K_{1}(\frac{%
\mid \mathbf{r}_{i}-\mathbf{r}_{j}\mid }{\lambda })\hat{\mathbf{r}}_{ij}$ is the vortex-vortex interaction and $%
\mathbf{F}_{i}^{vp}=-\sum_{k=1}^{N_{p}}F_{k}^{p}e^{-(\frac{|r_{i}-R_{k}|}{%
r_{p}})^{2}}({\mathbf{r}}_{i}-{\mathbf{R}}_{k})$ is the pinning attraction.
 Here, $\phi _{0}$
is the quantum of magnetic flux, $\lambda $ is the London penetration
length, $K_{1}$ is the special Bessel function and $f_{vv}$ is a
dimensionless parameter that can be related to the stiffness of the vortex
lattice. The $N_{p}$ pinning centers are located at random positions $%
\mathbf{R}_{k}$, with strength and range $F_{k}^{p}$ and $r_{p}$.We measure
lengths in units of $\lambda $, forces (per unit length) in units of $f_{0}=%
\frac{\phi _{0}^{2}}{8\pi ^{2}\lambda ^{3}}$, time in units of $t_{0}=\eta
\lambda /f_{0}$. We will consider $N_{v}=1600$, $L_{x}=40\lambda $, $L_{y}=%
\sqrt{3}L_{x}/2$, $N_{p}=25N_{v}$, $r_{p}=0.2\lambda $ and $F_{k}^{p}$
chosen from a Gaussian distribution with mean value $F^{p}=0.2$ and a
standard deviation of $0.01F^{p}$. The equations of motion are integrated
using a standard Euler algorithm with step $h=0.04t_{0}$, and a hard cut-off
$\Lambda =4\lambda $ for the vortex-vortex force. The
mean vortex velocity is defined as $\text{v}(t)=\frac{1}{N_{v}}%
\sum_{i}^{N_{v}}\text{v}_{i}(t)$, where $\text{v}_{i}(t)$ is the instantaneous
velocity of the i-vortex in the force direction. The observable used
to characterize the degree of order is the proportion of vortices
with 6 neighbors $P_{6}=1-n_{d}$, where $n_{d}$ is the
density of disclinations. $P_6$ is calculated at the end of each cycle.
 
\begin{figure}[t]
\includegraphics[width=86mm]{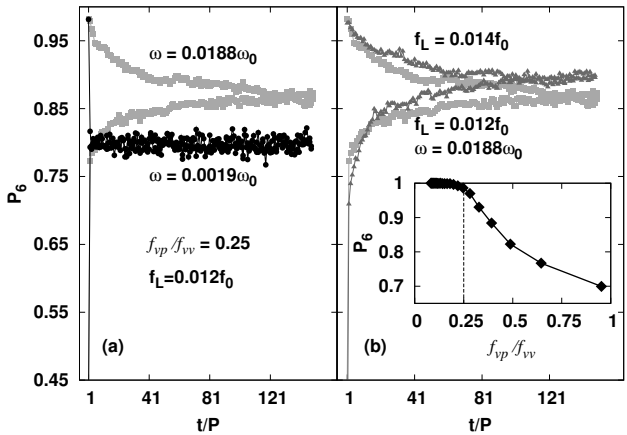}
\caption{ $P_{6}$ versus $t/P$ starting from an ordered and a disordered configuration.
(a) For two different $\omega$ and fixed $f_L$. (b) For two different $f_L$ and fixed $\omega$.
 (Inset) $P_{6}$ versus $f_{vp}/f_{vv}$,
showing a spontaneous creation of disclinations, and a decrease in $P_{6}$ at around $f_{vp}/f_{vv}=0.25$.}
\label{fig4}
\end{figure}

We assume that the main effect of temperature
near the PE is the increase of the ratio $f_{vp}/f_{vv}$, causing a
spontaneous disordering of the VL. In
the simulations, starting with a perfect ordered lattice (i.\thinspace e. $%
P_{6}=1$) and leading the system to evolve without any applied external
force, there is a spontaneous creation of disclinations, and a decrease in $%
P_{6}$.  In the inset of Fig. \ref{fig4}, the spontaneous $P_{6}$
starting from an ordered configuration is plotted as a function of $%
f_{vp}/f_{vv}$. We identify the parameters corresponding to the
experimental temperature region shown in Fig. \ref{fig1} as the range where
the VL spontaneously disorders. Therefore, we set the ratio $%
f_{vp}/f_{vv}=0.25$ to simulate the experimental system at a fixed
temperature (experiments of Fig. \ref{fig2}). The dc depinning force
per unit length $F_{c}^{dc}=\phi_{0} J_{c}    \approx  0.01f_{0}$ was
obtained by increasing a dc $F_{L}$ until the mean velocity $%
v$ exceeds a "criterion voltage". The shaking frequencies have been selected
in the low frequency regime ($\omega <<\omega _{c}\sim \frac{\alpha_L}{\eta }\sim 0.11$)
 \cite{diego}, where pinning forces dominate over viscous drags, because the experimental shaking  frequencies are well below $f_c = \omega_c / 2 \pi > 1 MHz $  

Figure \ref{fig3}(b) shows the calculated mean velocity as a function of
time during two shaking cycles for two different frequencies. After
switching the force direction, v decays towards a steady regime
characteristic of a dc driving force, similar to that reported in Ref. \onlinecite{reichhardt}
 that could be associated to a FSS. The transient time $%
\tau $ and the final mean velocity depend on the shaking force. At
very low frequencies, the system reaches the steady
regime inside each cycle (black curves in Fig. \ref{fig3}(b)) , and
the response is essentially dc. This is
not the case at higher frequencies (gray curve), were we observe that both $%
\tau $ and the final mean velocity slightly vary while the system
evolves from a cycle to the next one. This fact is more clear in
Fig. \ref{fig4}, that shows $P_{6}$ as a function of the number of
cycles of the shaking force, starting from different initial
conditions, at a fixed amplitude and two different frequencies
(\ref{fig4}(a)) and a fixed frequency and two amplitudes
slightly above the depinning force (\ref{fig4}(b)). After a\ new
transient time $\tau _{ac}$, that can involve many shaking cycles,
the disclination density remains fluctuating around a steady
value,\ in a dynamic state independent of the initial condition,
that depends on the amplitude of the shaking force. \ When the force
is very near the dc depinning force, a clear frequency dependence is
also observed (Fig. \ref{fig4}(a)).
 It becomes clear that the higher the shaking frequency
and amplitude of the shaking force (both effects expected by
increasing, in the experiment, the frequency of a shaking field), the more ordered the
resulting VLC, in agreement with that observed in our experiments.
In this process, there are 
two different transient times that involve the creation or annihilation
of disclinations: the characteristic dc time to decay to the FSS
 inside a cycle $\tau $ ( probably related with the
experimental transient time $\tau _{2}$ in Ref. \onlinecite{vortex2}\ ) 
and the transient time $\tau _{ac}$ that can take many shaking
cycles and could be related with the transient response observed   in ac
transport experiments \cite{vortex2, andrei}. This second process is inherent of an oscillatory drive, and dynamically
reorganizes the system in a VLC independent of the initial conditions.

In conclusion, a dynamic reorganization in the plastic regime that keeps memory of the frequency drive has been observed experimentally and numerically. Throughout the order disorder crossing phase boundary in $NbSe_2$ there is a large collection of VLCs, that can be accessed after the application of ac shaking fields of different frequency, either ordering or disordering the initial state. The order of the VLCs is determined by non invasive ac susceptibility measurements, showing that  higher  shaking frequencies lead to more ordered VLCs. The system forgets the initial condition and but keeps memory of the shaking frequency, suggesting that the nature of the final states is more complex than previously conjectured \cite{pasquini08}. Molecular dynamics simulations performed for a compatible set of interaction parameters in the plastic regime have reproduced qualitatively the salient features of experimental shaking protocols, revealing a plausible nature of the experimental VLCs. The dynamic steady states are reminiscent of FSS observed in other systems, as VL disclinations and mean vortex velocity fluctuate around a steady value. As a novel fact, we have identified two characteristic time constants: one characteristic of the dynamics inside each cycle (dc) and the second related to ac dynamics. Furthermore, as the theoretical model is similar to the one used in the simulations of other interacting systems \cite{reichhardt}, we expect that frequency dependent final states should be observed experimentally in a wide collection of particle systems with plastic flow regimes. A deeper study of the transient processes and its connection with the depinning transition will be object of future work. 

We thank E. Zeldov for useful discussions. This work was supported 
by UBACyT, ANPCyT and CONICET.


\begin{thebibliography}{99}
\bibitem{nota} As we are dealing with a low $T_{c}$ superconductor and creep
is negligible, a very large $n$ $>50$ has been set for $J<J_{c}$.
\end{thebibliography}
\end{document}